\documentclass[aps,twocolumn]{revtex4}
\usepackage[utf8]{inputenc}
\usepackage{amsmath,latexsym,amssymb,verbatim,enumerate,graphicx}
\usepackage{graphicx}

\newtheorem{lemma}{Lemma}

\newtheorem{thm}{Theorem}
\newtheorem{prop}{Proposition}

\newcommand{\tr}{{\rm Tr}}
\def\be{\begin{equation}}
\def\ee{\end{equation}}
\def\ben{\begin{eqnarray}}
\def\een{\end{eqnarray}}

\def\bei{\begin{itemize}}
\def\eei{\end{itemize}}

\def\ep{\epsilon}
\def\ecal{{\cal E}}

\DeclareUnicodeCharacter{981}{\phi}

\usepackage{color}

\begin{document}

\title{Bound on Bell Inequalities by Fraction of Determinism and \\ Reverse Triangle Inequality}

\author{ P. Joshi$^{1,2}$, K. Horodecki$^{1,3}$,  M. Horodecki$^{1,2}$, P. Horodecki$^4$, R. Horodecki$^1$, Ben Li$^5$ S. J. Szarek$^{5,6}$ and T. Szarek$^7$}
\affiliation{$^1$National Quantum Information Center of Gda\'nsk, 81--824 Sopot, Poland}
\affiliation{$^2$Institute of Theoretical Physics and Astrophysics, University of Gda\'nsk, Gda\'nsk, Poland}
\affiliation{$^3$Institute of Informatics, University of Gda\'nsk, 80--952 Gda\'nsk, Poland}
\affiliation{$^4$ Faculty of Applied Physics and Mathematics, Technical University of Gda\'nsk, 80--233 Gda\'nsk, Poland}
\affiliation{$^5$ Department of Mathematics, Case Western Reserve University, Cleveland, Ohio 44106-7058, U.S.A.}
\affiliation{$^6$ Institut de Math\'ematiques de Jussieu-PRG, Universit\'e Pierre et Marie Curie-Paris 6, 75252 Paris, France}
\affiliation{$^7$Institute of Mathematics, University of Gda\'nsk, 80--952 Gda\'nsk, Poland}

\date{\today}

\begin{abstract}
It is an established fact that entanglement is a resource. Sharing an entangled state leads 
to non-local correlations and to violations of Bell inequalities. 
Such non-local correlations illustrate the advantage of quantum resources over classical resources. 
Here, we study quantitatively  Bell inequalities with $2\times n$ inputs. 
As found in [N. Gisin et al., Int. J. Q. Inf. 5, 525 (2007)] 
quantum mechanical correlations cannot reach the algebraic bound for such inequalities. 
In this paper, we uncover the heart of this effect which we call the {\it fraction of determinism}.
We show that any quantum statistics with two parties and $2 \times n$ inputs exhibits nonzero fraction of determinism,
and we supply a quantitative bound for it. We then apply it to provide an 
explicit {\it universal upper bound} for Bell inequalities with $2\times n$ inputs. 
As our main mathematical tool we introduce and prove
a {\it reverse  triangle inequality}, stating in a quantitative way that if some states are  far away from a given state, 
then their mixture is also. The inequality is crucial in deriving the lower bound for the fraction of determinism, 
but is also of interest on its own.
\end{abstract}


\maketitle

\section{Introduction} \label{sec:intro}
Since Bell's paper \cite{Bell} entanglement has been studied and explored in depth. Saying that quantum information branch emerged from extensive studies of phenomenon of entanglement would not be an exaggeration. Entanglement has been used in many information-processing applications in which it either yields an advantage over the classical setting,  e.g., in communication complexity \cite{Buhrman-rev},  or where a classical counterpart simply doesn't exist, e.g., in quantum key distribution (QKD) \cite{Gisin-etal-QC-review}, its device independent variant  (DIQKD) \cite{acin-2007-98}, teleportation, super dense coding \cite{Nielsen-Chuang}, or Pseudo-Telepathy (PT) \cite{Brassard-pseudo, Bras_pt_survey}. 

Although quantum theory allows for 
violations of Bell inequalities (BI), in certain cases the violations can not reach their maximum algebraically possible value. Tsirelson was the first to find such upper bounds on Bell values for quantum theory \cite{Tsirelson-bound} and to relate  them to the Grothendieck's inequality. 
Much research has been done to explain why quantum mechanics does not lead to ``algebraic'' violations of Bell inequalities \cite{PR,ravi_qmnsbox}. In \cite{Oppenheim-Wehner}, Wehner and Oppenheim argue that the trade-off between steerability and uncertainty determines how non-local a theory is. In \cite{Cleve}, Cleve et al.  gave an upper bound for the winning probability for XOR games in the quantum setting;  their bound depends on the classical winning probability and the Grothendieck's constant. Note that XOR game is a non-local game and that non-local games form a subset of general Bell inequalities \cite{nonlocal_rev}. 

The approach to bounding quantum violations via a Grothendieck-type constant is now quite common and reasonably well understood. It leads to estimates for  Bell values that are of the form $\beta_{qm} \leq K_G \beta_{cl}$ \cite{junge_ubviol}. In this work we develop a different strategy, where the Bell value for a given inequality depends on the difference between the maximal algebraic value  ($\beta^{max}_{alg}$) and maximal deterministic value ($\beta^{max}_{det}$) of the inequality in question. 
Specifically, we study quantitatively Bell inequalities with $2\times n$ inputs (henceforth $2\times n$ BI)
and give a {\it universal bound} on quantum Bell values of these inequalities. To find this bound for $2\times n$ \,BI, we introduce notion of {\it fraction of determinism}  (FOD) and show that it depends only on the number of outcomes Alice and Bob have at their sites. We claim that presence of FOD prevents quantum Bell value from attaining the maximal algebraic value of a Bell type inequality. Our paper is inspired by Gisin et al. \cite{GisinMS2006-pseudo}, which studied
certain Bell inequalities (Pseudo-telepathy) 
for which quantum resources achieve algebraic violation. They show that to achieve such 
violations for these inequalities at minimum $3\times 3$ inputs are required. In other words, there is no $2\times n$  BI 
for which quantum theory attains algebraic violation. Here we uncover the heart of this effect -- the fraction of determinism -- and are able to give a quantitative bound for it.

\begin{figure}
    \centering
    \includegraphics[width=0.4\textwidth]{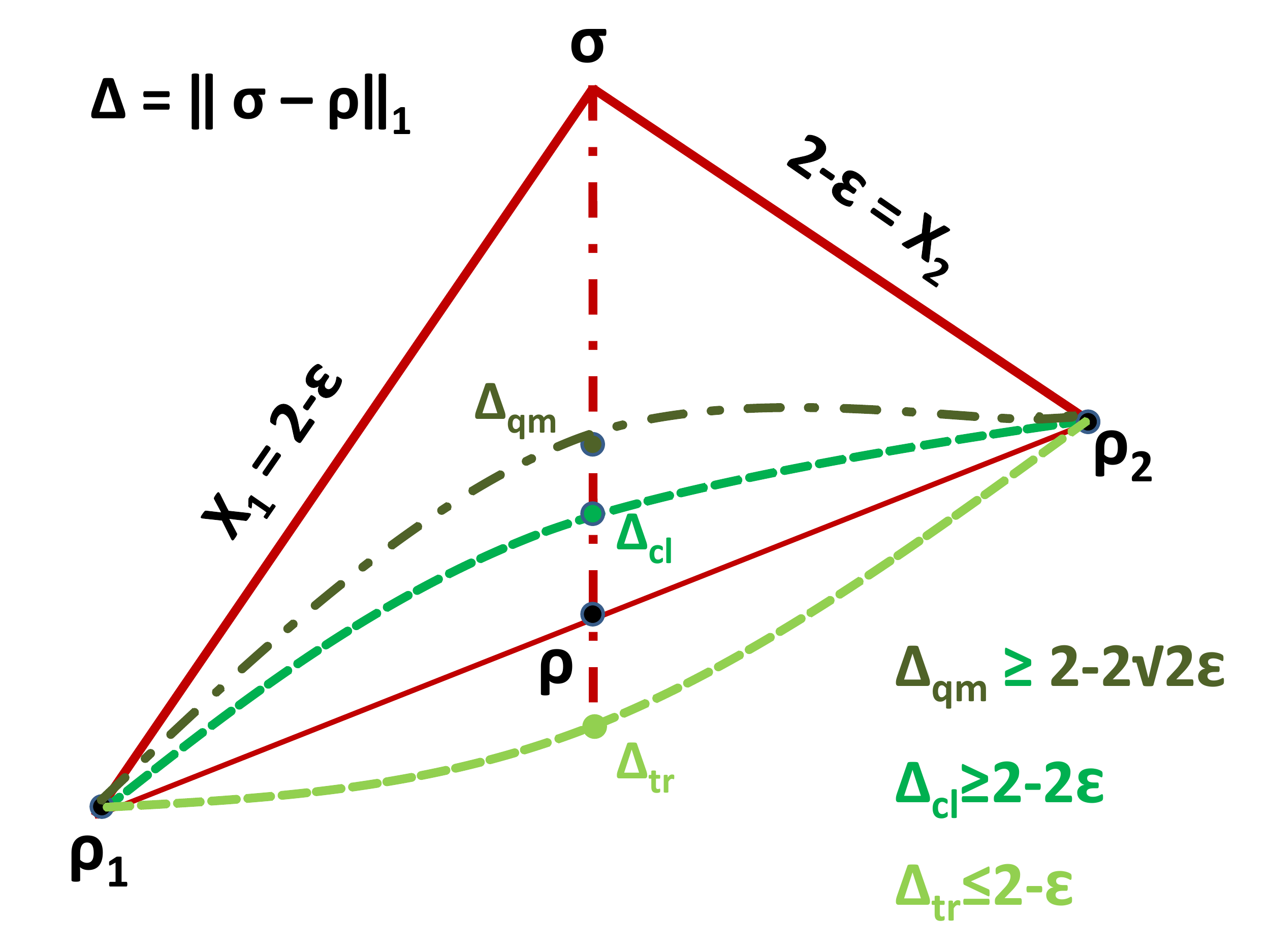} 
\caption{Pictorial representation of different bounds of $||\sigma-\rho||_1$. Triangle inequality gives an upper bound $2-\ep$, whereas {\it reverse triangle inequalities} give lower bounds $2-2\sqrt{2\ep}$ for general quantum states and  $2-2\ep$ for classical (or commuting) states.
\label{fig:trian_conj}}
\end{figure}

While looking for a lower bound for FOD, we proved a fundamental property of quantum states which is interesting on its own. Namely, if $\rho_1$ and $\rho_2$ are far from $\sigma$,  then any convex mixture of them is also far from $\sigma$. More precisely, if $\Delta_1=||\rho_1-\sigma||_1\ge 2-\epsilon$ and $\Delta_2=||\rho_2-\sigma||_1\ge 2-\epsilon$ for 
some $\epsilon \ge 0$, then, for all $p \in [0,1]$, 
\be
\Delta=||p\rho_1+(1-p)\rho_2-\sigma||_1\ge 2-\mathcal{O}(\sqrt{\epsilon}).
\ee
where $||\rho||_1 \overset{\text{def}}{=}Tr\sqrt{\rho^{\dagger}\rho}$. This inequality is in a sense dual to triangle inequality since it bounds the trace distance between $\rho$ and $\sigma$ from below. Accordingly, we call it ``reverse triangle inequality'' (RTI). Interestingly, it turns out that for classical states (commuting density matrices) one can find lower bound of $\Delta$ with the defect term linearly dependent on $\epsilon$, while for non-commuting quantum states one can not in general have dependence better than $\mathcal{O}(\sqrt{\epsilon})$.

The second fundamental property which is used here is related to so called steering \cite{wiseman_steering}. 
Namely, by making measurement on one site of the entangled state,  one can create only those ensembles which give rise to the same density matrix – the reduced state of the entangled state. 
This implies that if we consider two such ensembles, there must be at least two elements (one from one ensemble, and the other from the second ensemble that are not perfectly distinguishable. It has been apparently not studied to what extent they have to be indistinguishable. Here, by using the reverse triangle inequality, we are able to give a robust quantitative bound (lemma \ref{lemma-x}), which is independent of the dimension of the underlying Hilbert space. We shall use it further to give bound for FOD, which in turn will allow to bound quantum violations for $2\times n$ Bell inequalities.

The paper is organized in the following manner. In section \ref{sec:prelim}, we introduce necessary definitions and the concept of FOD. In sections \ref{sec:summary} and \ref{sec:FODinQM}, we present respectively a summary of our main results and sketches of their derivations. A special case when Bob has two inputs at his site with binary outcomes is analyzed in section \ref{sec:binary-bob}. For this special case, we have explicitly calculated a bound for FOD and for the classical fraction. Finally we conclude our work in section \ref{sec:conclusions}. 
Details of most proofs are relegated to the Appendices. 

\section{Preliminaries} \label{sec:prelim}
\subsection{Definitions} 
{\bf Box:} Consider two distant parties, Alice and Bob, sharing a physical system. Each of them perform measurements labeled as $x$ and $y$ respectively.  Their corresponding outcomes are labeled as $a$ and $b$. Then, a box is defined as family of joint probability distributions $p(a,b|x,y)$, i.e., $P=\{p(a,b|x,y)\}$. By a {\it non-signalling box} (NS-box) we mean a box which satisfies following conditions, 
\begin{flalign} 
p(b|y)=\sum_a p(a,b|x,y)=\sum_a p(a,b|x',y) \hspace{1mm} \forall b,x,x' \hspace{1mm} \text{and} \hspace{1mm}y \\ \nonumber
p(a|x)=\sum_b p(a,b|x,y)=\sum_b p(a,b|x,y') \hspace{1mm}\forall a,x,y' \hspace{1mm} \text{and} \hspace{1mm}y
\end{flalign}
A {\it local box} is defined as a box where joint probabilities can be expressed as 
\be
p(a,b|x,y)=\int_{\Lambda} q(\lambda)p(a|x,\lambda) p(b|y,\lambda) d\lambda , 
\ee
where the hidden variable $\lambda$ 
is distributed according to some probability density $q(\lambda)$. Such boxes satisfy (by definition, see below) every Bell inequality. 
We say that a box is a {\it Quantum box} (QM box) when conditional probabilities are realized by $p(a,b|x,y)=tr(M^x_a \otimes N^y_b \rho_{AB})$, where $\rho_{AB}$ is a shared quantum state between party A and B, and $M^x_a$ and $N^y_b$ are measurements for A and B respectively. Note that, for each input $x$ and $y$, $\{M^x_a\}, \{N^y_b\}$ are POVMs. In this work, we  consider only NS-boxes.

{\bf Bell Inequalities:}
Let $\mathcal{S}\equiv \{s^{a,b}_{x,y}\}$ be a real vector and $P=\{p(a,b|x,y)\}$ be a box then, $\mathcal{S}.P \leq \beta_T$ is called {\it Bell inequality} when this inequality is satisfied by any local box $P$ \cite{nonlocal_rev}. Note that we can rescale the inequality and make $\mathcal{S}$ positive real.

{\bf Fraction of determinism (FOD):}
Consider a non-signalling box $P$. One can always express it as convex combination of $P=(1-c)X+c D$, where $X$ is an NS-box and D is a deterministic box. The {\it fraction of determinism} of $P$ is defined as 
\begin{equation}
FOD:= \max_{D, X} \{c \,|\, P=(1-c)X+cD \}
\end{equation} 

{\bf Classical Fraction (CF):}
A non-signalling box $P$, can always be expressed as a convex combination of $P=(1-\sum_i c_i) X+ \sum_i c_i D_i$, where $X$ is an NS-box and $D_i$ are deterministic boxes. Let $c_{cf}=\sum_i c_i$ then the classical fraction of a box $P$ can be obtained by taking maximum of $c_{cf}$ over all possible decompositions of the above form, i.e.,
\be 
CF(P)= \max_{\{D_i\}, X} \{c_{cf}| P=(1-\sum_i c_i) X+ \sum_i c_i D_i\}
\ee
Note that FOD, CF and the cost of non locality $c_{nl}$ \cite{Brunneretal2011} satisfy the following relations
\be
FOD\leq CF=1-c_{nl}
\ee

\subsection{The Role of the Fraction of Determinism}
In the classical theory, FOD is not 1. Indeed, consider the maximally mixed state, which has the smallest fraction of determinism:   it is $1/k_A k_B$ where $k_A$ is number of Alice's outcomes, and $k_B$ is number of Bob's outcomes (assuming that the number of outcomes is same for all observables). In the quantum case, the set of states is larger, hence we might in principle have states with zero fraction of determinism. 
However this is not the case as shown here. On the other hand, PR-boxes \cite{PR} are completely noiseless and they do not have any fraction of determinism. The latter is equivalent to saying that they provide perfectly secure correlations. Indeed, the fraction of determinism is at the same time the fraction that can be known by the third party, or equivalently one can say that FOD in a given theory restricts Bell value from reaching its maximal algebraic value. We present the following proposition which indeed captures this idea. 

If a box has some fraction $c$ of "determinism'', then this fraction implies a bound on the maximal value of a linear function (in particular for Bell type inequalities).
\begin{prop}
\label{prop-c}
Consider a box $P=\{p(a,b|x,y)\}$ with inputs $x\in \{x_1,\ldots,x_n\}$ on Alice side and $y\in \{y_1,\ldots,y_m\}$ on Bob's side. Suppose that we can find indices $a^{(1)}_0, \ldots, a^{(n)}_0$, $b^{(1)}_0, \ldots, b^{(m)}_0$ such that 
\be
\forall i, j \hspace{5mm}
p(a^{(i)}_0,b^{(j)}_0|x_i,y_j)\geq c
\label{eq:cond}
\ee
Then for any linear function $\beta$ of the box, we have 
\be \label{eq:lin_func}
\beta(P) \leq \beta_{alg}^{\max} - c(\beta_{alg}^{\max} -\beta_{det}^{\max})
\ee
where $\beta_{alg}^{\max}$ is the maximum value over all boxes,
while $\beta_{det}^{\max}$ is the maximum over all classical deterministic boxes.
\end{prop}
This follows from the fact that any such box can be expressed as convex combination of a deterministic box and some other box, i.e., $P=cD+(1-c)X$ and simply taking the maximal value of $\beta$.

\section{Summary of The Results}  \label{sec:summary}
We give a {\it universal bound} on $2\times n$ Bell inequalities. 
Specifically, our main result is finding a bound on  FOD for the $2\times n$ BI scenario and showing that it only depends on number of outcomes of both parties. From Proposition \ref{prop-c} we know that this gives a universal bound for any linear function. A summary of our main results is as follows. 

\begin{thm}\label{thm:main}
For input $2\times n$, the fraction of determinism for QM box is bounded by the following quantity: 
\be 
c\geq {0.1134\over 2k\hspace{1mm} l \hspace{1mm}l_1\hspace{1mm} l_2}
\ee
Here, $k=\max\{|x_1|,...|x_n|\}$ and $l=\max\{l_1=|y|,l_2=|y'|\}$, where $\{x_1,...x_n\}$ are inputs on Alice's side while $\{y, y'\}$ are inputs on Bob's side.
\end{thm}
Here, by $|z|$ we denote number of outcomes an observable $z$ takes.

To prove the above  theorem we need the following fundamental property of quantum states. 
\begin{thm}
\label{thm:szarek}
{\bf (Reverse Triangle Inequality)}Let $\ep \geq 0$ and assume that the states $\rho_i, \sigma$ satisfy 
\be \label{eq:condtn}
\|\rho_i -\sigma\|\geq 2 - \ep
\ee
for  $i=1,\ldots,l$. Then, for any probability distribution $\{p_i\}_{i=1}^l$,
\bei
\item[1)] For any states $\rho_i$, $\sigma$ satisfying \eqref{eq:condtn}
 \be \label{eq:szarek}
 \|\sum_{i=1}^l p_i \rho_i - \sigma\| \geq 2 - 2\sqrt{l\ep}
\ee
\item[2)] For commuting states $\rho_i$, $\sigma$ satisfying \eqref{eq:condtn}
\be
 \|\sum_{i=1}^l p_i \rho_i - \sigma\| \geq 2 - l\ep
\ee
\item[3)] There exist three non-commuting states $\rho_1, \rho_2$ and $\sigma$ satisfying \eqref{eq:condtn} such that
\be
\| {\rho_1+ \rho_2 \over 2}  - \sigma\| \leq 2-\sqrt{2\ep}
\ee 
\eei
\end{thm}
{\bf Remark}: The third assertion says that, in the non-commuting case, $2-\sqrt{2\ep}$ is the best possible bound one can hope to achieve. Hence, one cannot have better lower bound than $2-\mathcal{O}(\sqrt{\ep})$.

Using the above results, one can find lower bound for FOD in CHSH \cite{nonlocal_rev} case ($k=l=2,\beta_{alg}^{\max}=4$ and $\beta_{det}^{\max}=2$) as $c \geq 3.5438 \times 10^{-3}$. This results in bounding CHSH value for quantum theory.
\begin{flalign} \label{beta_thm} \nonumber
\beta_{CHSH}^{qm} &\leq \beta_{alg}^{\max} - c(\beta_{alg}^{\max} -\beta_{det}^{\max}) &\\
&\leq 4-3.5438 \times 10^{-3}(4-2)=3.9929
\end{flalign}

A more direct approach gives an improved bound on FOD.
\be
\beta_{FOD}(P)\le 4-\frac{0.1096*2}{4}=3.9452
\ee
\be
\beta_{CF}(P)\le 4-\frac{0.1123*2}{4}=3.9439
\ee
This has been elaborated in section \ref{sec:binary-bob}.

It is interesting to note that we can also roughly estimate $\beta$ of \eqref{eq:lin_func} to upper-bound $\beta$ in the classical theory. We will get a rough estimation for CHSH (in case of the maximally mixed state $c={1\over k_A k_B}={1\over4}$) 
\be
\beta_{CHSH}^{cl} \leq 4-{1\over 4}(4-2)=3{1\over 2}
\ee

We realize that these are weak bounds, but the importance of this study lies in their generality: they are valid for {\em any} Bell inequality.
In the following section we shall find a bound for $c$ for quantum states for quantum theory and derive our main results. Most of the proofs are relegated to Appendix \ref{app:FODinQM}. We assume that Bob has 2 observables $\{y,y'\}$, i.e., $m=2$.

\section{Fraction of Determinism in QM} \label{sec:FODinQM}
We start with a proposition in which we redefine FOD more explicitly for QM boxes, which will lead to a lower bound that can be used in Proposition {\ref{prop-c}}. 
\begin{prop}
For a QM-box with $2\times n$ input, the following quantity $c_0$ satisfies \eqref{eq:cond}
\be \label{classicbound}
c_0=\inf_{\xi,\xi',\{X_r\}} \max_{r,i,j} \min\{ p_i \tr(X_r \rho_i), q_j\tr(X_r \sigma_j)\}
\ee
where the infimum is taken over all ensembles $\xi=\{(p_i,\rho_i)\}_{i=1}^{|y|}$,
$\xi'=\{(q_j,\sigma_j)\}_{j=1}^{|y'|}$ satisfying
\be \label{eq:ensemble}
\sum_ip_i \rho_i = \sum_j q_j \sigma_j
\ee
and over all POVMs $\{X_r\}_{r=1}^k$, i.e., sets of operators satisfying $\sum_r X_r=I$, $X_r\geq 0$,
with $k=\max\{|x_1|,\ldots,|x_n|\}$.
\end{prop}
 
{\bf Proof}: By hypothesis, our quantum box is realized via 
 POVMs $\{M^x_a\}$ (with $x\in \{x_1,\ldots,x_n\}$) on  on Alice's side, two  POVMs  $\{N^y_b, N^{y'}_{b'}\}$ on Bob's side, and a shared quantum state $\rho_{AB}$. 
 Depending on Bob's measurement choice ($y$ or $y'$), an ensemble  $\{p(b|y),\rho_b\}_{b=1}^{|y|}$ or $\{p(b'|y'),\sigma_{b'}\}_{b'=1}^{|y'|}$  is created at Alice's site, where $p(b|y)$ and $p(b'|y')$ are marginal conditional probabilities. 
 Even more specifically, $p(b|y)\rho_b = \tr_B\big((I\otimes N^y_b) \rho_{AB}\big)$ and similarly $p(b'|y')\sigma_{b'} = \tr_B\big((I\otimes N^{y'}_{b'}) \rho_{AB}\big)$. These ensembles satisfy 
 \be 
 \tr_B(\rho_{AB})=\sum_b p(b|y)\rho_b =\sum_{b'} p(b'|y')\sigma_{b'}  ,
 \ee 
 i.e., a condition of the type \eqref{eq:ensemble}.  
 If now $\{X_r\}$ is any of Alice's POVMs (say, $\{M^x_a\}$), it is apparent that the expressions $p_i \tr(X_r \rho_i), q_j\tr(X_r \sigma_j)$ 
 coincide with the conditional probabilities $p(a,b|x,y), p(a,b'|x,y')$ appearing in \eqref{eq:cond}. 
 Now pick a triplet $(r, i, j)$ such that the probabilities of the corresponding outcomes are maximal and 
 one can see that these indices lead to the choices of $a, b$ that yield \eqref{eq:cond}  with $c=c_0$.
 $\blacksquare$

Next, we will give an estimate on this quantity. In this way we shall obtain a universal quantum bound for any $2\times n$ input inequalities, in terms of difference between the classical bound and the maximal algebraic bound (\ref{eq:lin_func}). In general, $c$ might be zero. But we show in $2\times n$ input boxes that indeed it is bounded away from zero. To show this, one needs to prove for some choice of $i,j$ and for any POVM $X_r$, that $\tr(X_r \rho_i)$ and $\tr(X_r \sigma_j)$ are bounded away from zero. Note that this indeed happens when the POVM cannot distinguish the two states $\rho_{i}$  and $\sigma_j$ too well. We elaborate this point through the following lemma.

\begin{lemma}
Suppose that $||\rho-\sigma||\leq 2- 2 k \ep$. 
Then for any POVM $\{X_r\}_{r=1}^k$ there exists an outcome $r_0$ such that 
\be
\tr (X_{r_0} \rho) \geq \ep \quad \hbox{and} \quad \tr (X_{r_0} \sigma) \geq \ep
\label{eq:error}
\ee
\label{lem:error}
\end{lemma}

Note that using this lemma we can replace conditional probabilities by $\epsilon$ and get rid of choosing maximum for all $r$ and the optimization over $\{X_r\}$. The above lemma asserts that there exist at least one outcome $r$ for each input and each pair $(\rho_i,\sigma_j)$ such that the corresponding probabilities are lower-bounded by $\epsilon_{ij}$, i.e., $\tr (X_{r} \rho_i) \geq \ep_{ij}$ and $\tr (X_{r} \sigma_j) \geq \ep_{ij}$. Therefore one can simplify the expression for FOD as follows. 
\begin{align} \nonumber
&c_0\ge c_1=\inf_{\xi,\xi',\{X_r\}} \max_{r,i,j} \min\{ p_i \epsilon_{ij}, q_j\epsilon_{ij}\} &\\ 
&c_1\ge \frac{1}{2k} \inf_{\xi,\xi'}\max_{i,j} \min\{p_i (2- ||\rho_i-\sigma_j||), q_j (2- ||\rho_i-\sigma_j||)\}&
\label{eq:c1}
\end{align}
where we assume $||\rho_i-\sigma_j||\leq 2- 2 k \ep_{ij}$.

Having simplified FOD, we will now state and apply a theorem which is both vital for our results, as well as important on its own. 

\addtocounter{thm}{-1}
\begin{thm}
\label{thm:szarek}
\{Restatement\}Let $\ep \geq 0$ and assume that the states $\rho_i, \sigma$ satisfy 
\be \label{eq:condtn1}
\|\rho_i -\sigma\|\geq 2 - \ep
\ee
for  $i=1,\ldots,l$. Then, for any probability distribution $\{p_i\}_{i=1}^l$,
\bei
\item[1)] For any states $\rho_i$, $\sigma$ satisfying \eqref{eq:condtn1}
 \be \label{eq:szarek}
 \|\sum_{i=1}^l p_i \rho_i - \sigma\| \geq 2 - 2\sqrt{l\ep}
\ee
\item[2)] For commuting states $\rho_i$, $\sigma$ satisfying \eqref{eq:condtn1}
\be
 \|\sum_{i=1}^l p_i \rho_i - \sigma\| \geq 2 - l\ep
\ee
\item[3)] There exist three non-commuting states $\rho_1, \rho_2$ and $\sigma$ satisfying \eqref{eq:condtn1} such that
\be
\| {\rho_1+ \rho_2 \over 2}  - \sigma\| \leq 2-\sqrt{2\ep}
\ee 
\eei
\end{thm}

We relegate the proof of the above Theorem to Appendix \ref{app:fid_prof}. 
 
Using this theorem we argue that for two ensembles (\ref{eq:ensemble}), which give rise to the same density matrix, $||\sigma_{i_0}-\rho_{j_0}||$ must be bounded away from 2 for some $i_0,j_0$. In general, we have the following lemma.

\begin{lemma}\label{lemma-x}
For two ensembles $\{(p_i,\rho_i)\}_{i=1}^{|y|}$,
$\{(q_j,\sigma_j)\}_{j=1}^{|y'|}$ satisfying 
\be
||\sum_ip_i \rho_i - \sum_j q_j \sigma_j||\leq x
\ee
there exist $i_0$ and $j_0$ such that 
\be
||\rho_{i_0}-\sigma_{j_0}||\leq 2 - \ep
\ee
where $\ep$ is solution of the following equation
\be\label{eq-x}
2 - 2\sqrt{l_1l_2\ep}=x
\ee
where $|y|=l_1$ and $|y'|=l_2$.
\end{lemma}

We are now almost done. However,  it may still happen that, for the chosen pair of indices, 
the probabilities $p_{i_0}$,  $q_{j_0}$ are small,
and we will not have a bound for the whole quantity of \eqref{eq:c1}.
Therefore we need to truncate the ensembles so that the minimal probability 
is bounded away from zero. Such smaller ensembles, 
do not give rise anymore to the same density matrix. 
However their density matrices are still close, provided we have not truncated too much. 

\begin{lemma}
Suppose we are given two ensembles 
\be
\ecal_1= \{p_i,\rho_i\}_{i=1}^{l_1},\quad \ecal_2= \{q_j,\sigma_j\}_{j=1}^{l_2}
\label{eq:ens2}
\ee
which give rise to the same density matrix. Let $p_i$ and $q_j$ 
be arranged in the nonincreasing order.  Let us denote
\be
\delta_1= 1- \sum_{i=1}^{\tilde{l_1}} p_i,\quad \delta_2= 1- \sum_{j=1}^{\tilde{l_2}} q_j
\ee
Consider new ensembles 
\be
\tilde\ecal_1= \{\tilde p_i,\rho_i\}_{i=1}^{\tilde{l_1}},\quad \tilde\ecal_2= \{\tilde q_j,\sigma_j\}_{j=1}^{\tilde{l_2}}
\label{eq:ens2}
\ee
where $\tilde p_i = p_i/(1-\delta_1)$,$\tilde q_j = q_j/(1-\delta_2)$.
Then the new ensembles satisfy
\be
\|\sum_{i=1}^{\tilde{l_1}}\tilde p_i \rho_i - \sum_{j=1}^{\tilde{l_2}}\tilde q_j \sigma_j\| \leq \frac{2 \max\{\delta_1,\delta_2\}}{1 - \min\{\delta_1,\delta_2\}}
\ee
\label{lem:cut}
\end{lemma} 

Thus we can use the new ensembles to show that there exist a pair of states $\rho_{i_0}$ and $\sigma_{j_0}$, and that at the same time the weights of the states 
satisfy $p_{i_0} \geq p_{\tilde{l_1}}$, $q_{j_0} \geq q_{\tilde{l_2}}$. Thus adjusting $\tilde{l_1}$ and $\tilde{l_2}$ properly we can simultaneously secure a bound on both the weights and the norm.

\medskip We can now prove our final result.
\addtocounter{thm}{-2}
\begin{thm}\label{thm:main}
\{{\it Restatement}\} For input $2\times n$, the fraction of determinism for QM box is bounded by the following quantity 
\be \nonumber
c\geq {0.1134\over 2k\hspace{1mm} l \hspace{1mm}l_1\hspace{1mm} l_2}
\ee
Here, $k=\max\{|x_1|,...|x_n|\}$ and  $l=\max\{l_1=|y|,l_2=|y'|\}$, where $\{x_1,...x_n\}$ are inputs on Alice's side while $\{y, y'\}$ are inputs on Bob's side.
\end{thm}

{\bf Proof.} First truncate the ensembles appropriately. 
We use the notation of Lemma \ref{lem:cut}. Let $\mu>1$ be a parameter. Let us choose the largest $\tilde{l_1}$ and $\tilde{l_2}$ such that $p_{\tilde{l_1}} > \frac{1}{l\mu}$ and $q_{\tilde{l_2}}>\frac{1}{l\mu}$, where $l=\max\{l_1,l_2\}$. Then  $\delta_1$ and $\delta_2$ not larger than $l \times\frac{1}{l\mu}$.
Consequently, we get the following estimate on the truncated ensemble: 
\be
\|\sum_{i=1}^{\tilde{l_1}} \tilde p_i \rho_i - \sum_{j=1}^{\tilde{l_2}}\tilde q_j \sigma_j\| \leq \frac{2/\mu}{1-1/\mu}= \frac{2}{\mu-1}.
\ee
From equation \eqref{eq:c1} and from Lemma  \ref{lemma-x} it follows that
\be
c\ge\frac{1}{2k}\frac{\ep}{l\mu},
\ee
where $\ep$ satisfies equation (\ref{eq-x}) with $x=\frac{2}{\mu-1}$. After some simplifications, we get
\be \label{eq:ep_sol}
\ep\ge \left(\frac{(\mu-2)}{l(\mu-1)}\right)^{2}.
\ee
Substituting this $\ep$ value in the preceding  equation we are led to 
\be
c\geq {1\over 2kl^3} \left({(\mu-2)\over (\mu-1)}\right)^2{1\over \mu} .
\ee
We now note that the function $f(\mu)=\frac{1}{\mu}\left(\frac{\mu-2}{\mu-1}\right)^2$ reaches its maximum at $\mu_0=(5+\sqrt{17})/2$, 
which completes the proof.
$\blacksquare$

{\bf Example}: Consider the CHSH case,  where $k=2, l=2$ and substituting these values we find 
$FOD \ge 7.0875 \times 10^{-3}$. Consequently 
\be
\beta_{CHSH}^{qm} \le \beta_{alg}-c (\beta_{alg}-\beta_{det})=3.9858.
\ee

In the next section bounds for FOD and CF are calculated for a simple case of $2\times n$ input  with binary outcomes on Bob's side. One can find these bounds using some of the lemmas and propositions described in section \ref{sec:summary}, which in turn gives an even better bound than the ones obtained using the general result of Theorem \ref{thm:main}.

\section{FOD and CF for Binary outcomes on Bob's side} \label{sec:binary-bob}
\begin{figure}
\begin{center}
\includegraphics[width=0.42\textwidth]{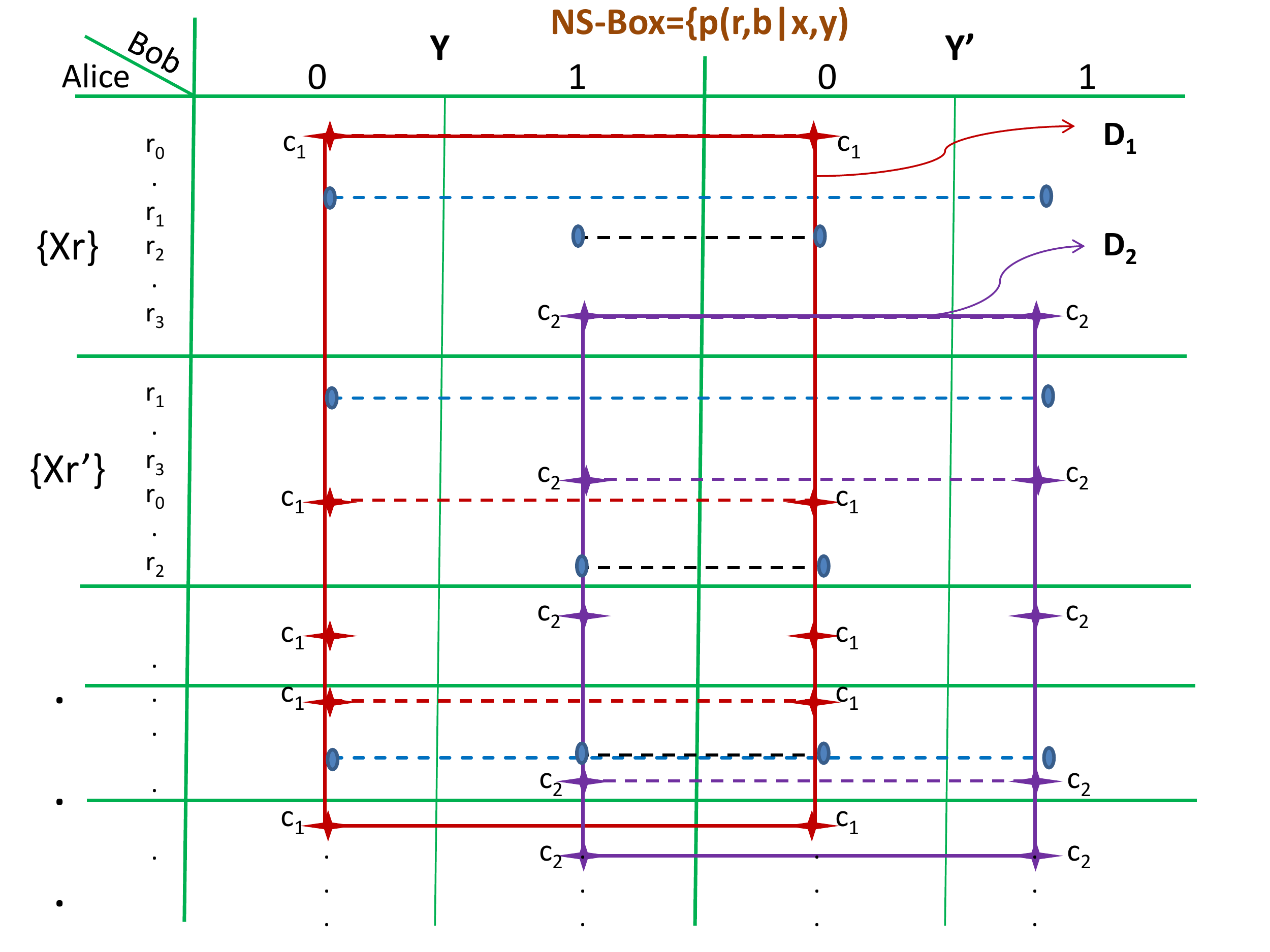}
\end{center}
\caption{The box $P=\{p(r,b|X_r, y)\}$ of Alice and Bob. $D_1$ and $D_2$ are two orthogonal deterministic boxes with fraction $c_1$ and $c_2$ respectively. And these can be subtracted from $P$. 
}
\label{det_box}
\end{figure}

Using structural property of boxes and Lemmas \ref{lem:error} and \ref{lem:cut} and Theorem \ref{thm:szarek} in section \ref{sec:FODinQM}, one can explicitly compute bounds for FOD and CF for the case when Bob has binary outcomes. Technically, we look for structures of deterministic boxes within the structure of the quantum box. The maximum fraction of these deterministic boxes bound FOD of the quantum box. This technique is explained below and on Fig. {\ref{det_box}}.

Bob can create $\{p_i\rho_i\}^1_{i=0}$ or $\{q_j\sigma_j\}^1_{j=0}$ ensemble at Alice's site by making measurement $y$ or $y'$ respectively on his part of shared quantum state. Lemma \ref{lem:error} asserts that for all pairs of $\rho_i$ and $\sigma_j$ and for all POVMs $\{X_r\}$
\begin{flalign} \label{eqn:confuse}
\exists \hspace{5mm} \epsilon_{ij} \ge 0, X_{r_0}, X_{r_1}, X_{r_2}, X_{r_3} \hspace{3mm}\hbox{such that} \\ \nonumber
tr(X_{r_0}\rho_0)\ge \epsilon_{00} , \hspace{3mm}{\text and} \hspace{3mm} tr(X_{r_0}\sigma_0)\ge \epsilon_{00} \\ \nonumber
tr(X_{r_1}\rho_0)\ge \epsilon_{01} , \hspace{3mm}{\text and} \hspace{3mm} tr(X_{r_1}\sigma_1)\ge \epsilon_{01} \\ \nonumber
tr(X_{r_2}\rho_1)\ge \epsilon_{10} , \hspace{3mm}{\text and} \hspace{3mm} tr(X_{r_2}\sigma_0)\ge \epsilon_{10} \\ \nonumber
tr(X_{r_3}\rho_1)\ge \epsilon_{11} , \hspace{3mm}{\text and} \hspace{3mm} tr(X_{r_3}\sigma_1)\ge \epsilon_{11}\ ,
\end{flalign}
where $\epsilon_{ij} \le {1\over 2k}(2-||\rho_i-\sigma_j||) $. This means that when Bob obtains outcomes $(b,b')$ for inputs $(y,y')$ then for any POVM of Alice there exist at least one outcome, call it a {\it confusing outcome}, on her side such that once she obtains it, she cannot distinguish between measurement choices of Bob with certainty, i.e., to determine whether Bob chose $y$ or $y'$ to create the first ensemble. For example, in the first pair of inequalities in \eqref{eqn:confuse} above, the outcome $r_0$ of some POVM can not tell apart with certainty $\rho_{00}$ from $\sigma_{00}$. There are four pairs of $(b, b')$, hence there are four confusing outcomes corresponding to each of these four cases.

Consider the particular case when Bob obtains $(0,0)$ when he measures $(y,y')$, and let us say $r_0$ is a confusing outcome for Alice when she chooses to measure POVMs $\{X_r\}$. Since Bob obtains $(0,0)$, the marginals satisfy $p_0 > 0$ and $q_0>0$. Lemma \ref{lem:error} asserts that for any measurement choice we have $\tr(X_{r_0}\rho_0)\ge \epsilon_{00}$ and $\tr(X_{r_0}\sigma_0)\ge \epsilon_{00}$. Hence for every POVM, there is at least one confusing outcome on Alice's side. Therefore, in the quantum box we can replace the probabilities corresponding to each of these confusing outcomes for every measurement choice of Alice with $c_{00}:= \min\{p_0\epsilon_{00}, q_0\epsilon_{00}\}$. One can now see that by this construction we can create a deterministic box (say $D_{00}$) with fraction equal to $c_{00}$. In other words, every quantum box $P_Q$ satisfies the relation $P_Q\ge (1-c_{00})X+c_{00}D_{00}$. In such a way, we can create four separate deterministic boxes ($\{D_{ij}\}^1_{i,j=0}$) corresponding to each of the outcome pairs $(b,b')$ of Bob.

\begin{figure}
\begin{center}
\includegraphics[width=0.42\textwidth]{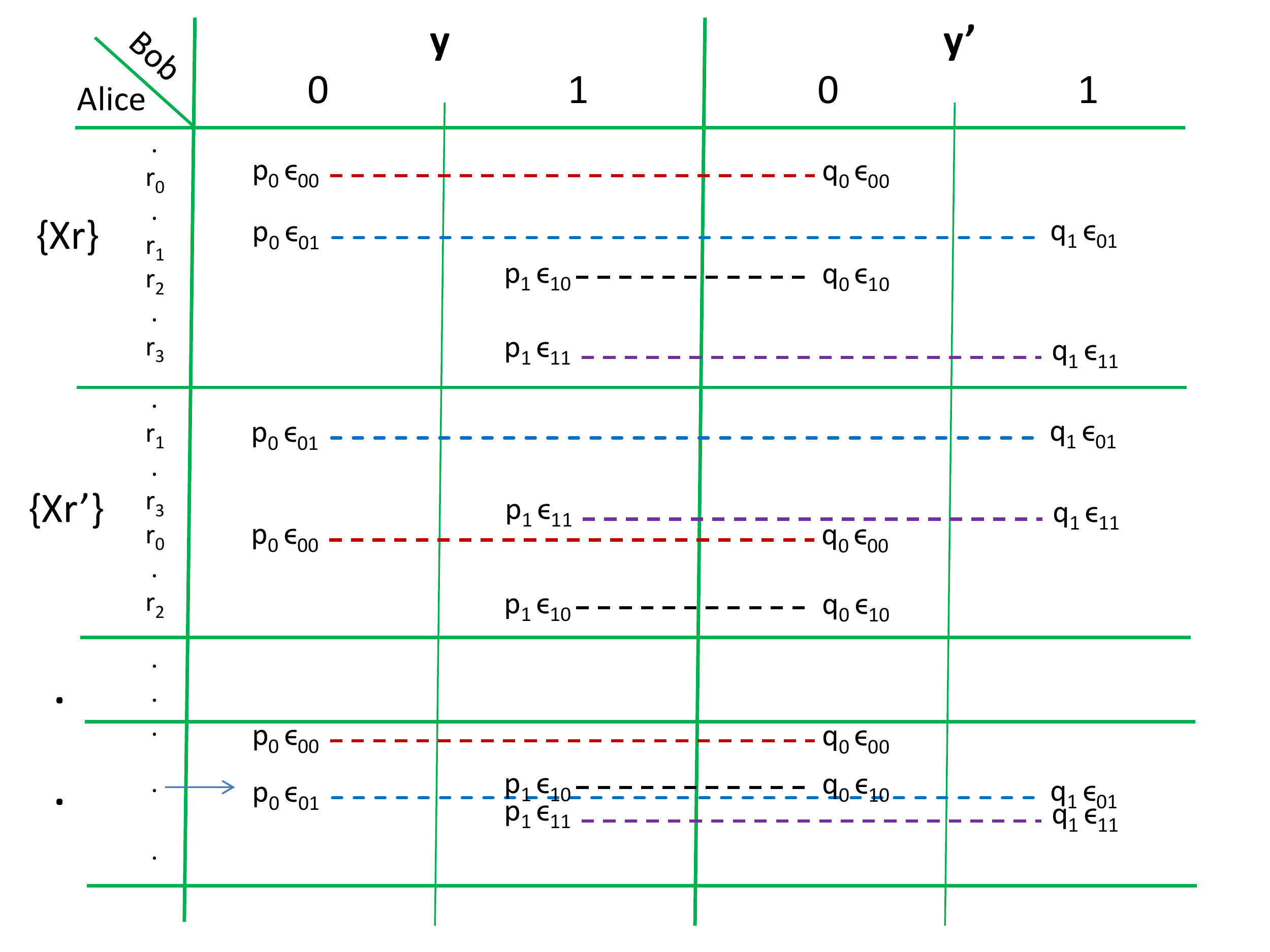}
\end{center}
\caption{The box $\{p(r,b|X_r, y)\}$ of Alice and Bob. Dashed lines represents which pairs are being confused and their lower bounds. Note that $r_i$'s are independent of which input Alice chooses. Some or all $r_i$'s may coincide with each other for some inputs of Alice.}
\label{clas_frac}
\end{figure}

There is a possibility that there may exist a measurement setting for Alice such that she obtains a single confusing outcome for two or more different cases, e.g., when she obtains a confusing outcome $r_0$, she is unable to distinguish between measurement choices of Bob not only in the case when Bob obtains $(0, 0)$ but also in the case when he obtains $(1, 0)$. So, in the worst case, for some measurement choices there may be just one confusing outcome at Alice's side for all the four different cases as in Fig.  \ref{clas_frac} in the last row of the box. In that case, the quantum box does not satisfy $P_Q\ge (1-\sum c_{ij})X+\sum^1_{i,j=0} c_{ij} D_{ij}$ because this would require us to use some probabilities twice. Therefore, in general ,one can use only orthogonal pairs of deterministic boxes to resolve this issue, i.e., either  $P_Q\ge (1-c_{00}-c_{11})X+c_{00} D_{00}+c_{11} D_{11}$ or $P_Q\ge (1-c_{01}-c_{10})X+c_{01} D_{01}+c_{10} D_{10}$.

The maximum fraction of such deterministic boxes bounds from below the FOD of the QM box under consideration. The sum of these fractions bounds CF. So to calculate FOD and CF for a fixed ensemble, we need to find
\begin{flalign}
FOD=\frac{1}{2k}\max\{&\min\{p_0 \epsilon_{00},q_0 \epsilon_{00}\},\min\{p_1 \epsilon_{11}, q_1 \epsilon_{11}\}, &\\ \nonumber
&\min\{p_0 \epsilon_{01},q_1 \epsilon_{01}\},\min\{p_1 \epsilon_{10},q_0 \epsilon_{10}\}\}&
\end{flalign}
and
\begin{flalign}
CF=\frac{1}{2k}\max\{&\min\{p_0 \epsilon_{00},q_0 \epsilon_{00}\}+\min\{p_1 \epsilon_{11}, q_1 \epsilon_{11}\}, &\\ \nonumber
&\min\{p_0 \epsilon_{01},q_1 \epsilon_{01}\}+\min\{p_1 \epsilon_{10},q_0 \epsilon_{10}\}\}&
\end{flalign}
 
To calculate these values w.l.g. we can assume $p_0\le q_0\le q_1\le p_1$.
Using lemma(\ref{lem:cut}) and optimizing over p's and q's we finally get the following values (appendix contains detailed calculations) \\
Using theorem \ref{thm:szarek} we find, $FOD=\frac{0.1096}{2k}$,  $CF=0.1122/2k$ and for k=2
\be
\beta_{FOD}(P)\le 4-\frac{0.1096*2}{4}=3.9452
\ee
\be
\beta_{CF}(P)\le 4-\frac{0.1123*2}{4}=3.9439
\ee
\\
These bounds are very weak, but since they hold for any $2\times n$ Bell type inequalities, 
they presumably can not be much better than this.

\section{Conclusion} \label{sec:conclusions}
Here we have given quantitatively a {\it universal bound} for $2\times n$ input Bell inequalities, which is independent of the number `n' of inputs. Specifically, we show that this universal bound depends on the number of outputs of the two parties and on the difference between the maximal algebraic value and the maximal deterministic value of the inequality. We show that presence of FOD in $2\times n$ BI prevents quantum Bell values from achieving the maximal algebraic value. Hence this result is also a quantitative proof of the theorem shown by Gisin et al. in \cite{GisinMS2006-pseudo}, which states that there exist no $2\times n$ input Pseudo-Telepathy game. Although these bounds are not tight, one can improve them by considering the classical fraction and generalize the result using it. We have analyzed a simple case where the classical fraction gives better bound than taking into account merely FOD.

To obtain the above results, we established a {\it reverse triangle inequality}, which is an independent result of its own interest.  The triangle inequality gives upper bounds on trace distance between two states, whereas RTI bounds the trace distance from below. We have determined that this bound is different for non-commuting states than when considering only commuting states. The bound in the commuting case is sharp, and the one in the non-commuting case is close to being sharp.

\begin{acknowledgments}
We thank Aram Harrow for a suggestion that lead to a simpler proof of our geometric result and to a  better constant. PJ thanks P. Mazurek for useful discussions. This work was supported by ERC QOLAPS, EC IP QESSENCE, EC grant RAQUEL, MNiSW grant IdP2011 000361 and NCBiR-CHIST-ERA Project QUASAR.  PJ was also supported by grant MPD/2009-3/4 from Foundation for Polish Science. SJS was partially supported by grants from the National Science Foundation (U.S.A.)  and by the grant 2011-BS01-008-02 from ANR (France).
TS was partially supported by the National Science Centre of Poland, grant number DEC-2012/07/B/ST1/03320. Part of this work was done at the National Quantum Information Centre of Gda{\'n}sk. 
\end{acknowledgments}

%

\clearpage

\appendix

\addtocounter{thm}{0}
\addtocounter{prop}{-1}
\addtocounter{lemma}{-3}
\addtocounter{cor}{-1}

\section{Proof of Theorem 2 and a discussion of its optimality} \label{app:fid_prof} 
\begin{thm}
\label{thm:szarek}
\{Restatement\}Let $\ep \geq 0$ and assume that the states $\rho_i, \sigma$ satisfy 
\be \label{eq:condtn2}
\|\rho_i -\sigma\|\geq 2 - \ep
\ee
for  $i=1,\ldots,l$. Then, for any probability distribution $\{p_i\}_{i=1}^l$,
\bei
\item[1)] For any states $\rho_i$, $\sigma$ satisfying \eqref{eq:condtn2}
 \be \label{eq:szarek}
 \|\sum_{i=1}^l p_i \rho_i - \sigma\| \geq 2 - 2\sqrt{l\ep}
\ee
\item[2)] For commuting states $\rho_i$, $\sigma$ satisfying \eqref{eq:condtn2}
\be
 \|\sum_{i=1}^l p_i \rho_i - \sigma\| \geq 2 - l\ep
\ee
\item[3)] There exist three non-commuting states $\rho_1, \rho_2$ and $\sigma$ satisfying \eqref{eq:condtn2} such that
\be
\| {\rho_1+ \rho_2 \over 2}  - \sigma\| \leq 2-\sqrt{2\ep}
\ee 
\eei
\end{thm}

\noindent {\it Proof}: We start by recalling two well-known facts. 

\smallskip \noindent
{\bf Rotfel'd Inequality} \cite{bourin}: Let $f$ be a concave function on $[0,\infty)$ such that $f(0)\geq0$ and let  $A_1, \ldots,A_l \geq 0$. Then 
\be \label{trac_inq}
Tr f\Big(\sum_{i=1}^l A_i \Big) \leq Tr \sum_{i=1}^l f(A_i)
\ee
Rotfel'd Inequality is usually stated for just two matrices (i.e., $l=2$), but the general case 
follows easily by induction. 

\smallskip \noindent
{\bf Fuchs–van de Graaf inequalities} \cite{Fuchs-Graaf}: These inequalities give two-sided 
bounds for the trace distance between two quantum states $\sigma$ and $\tau$  in terms of {\em fidelity} between $\sigma$ and  $\tau$, which is defined as $F(\sigma, \tau)=Tr\sqrt{\sqrt{\sigma} \, \tau \sqrt{\sigma}}$. We have 
\be \label{eqn:graaf_inq}
1-F(\sigma, \tau)\leq {1\over2}\|\sigma-\tau\|\leq\sqrt{1-F(\sigma, \tau)^2} .
\ee

 Rotfel'd inequality applied with $f(t) = \sqrt{t}$ allows us to upper-bound fidelity of the mixture 
 $\sum_{i=1}^l p_i \rho_i =: \rho$  in terms of individual fidelities: 
\ben \label{eqn:fid}
F(\sigma,\rho)&
=&Tr\sqrt{\sqrt{\sigma}\Big(\sum_{i=1}^l p_i \rho_i \Big)\sqrt{\sigma}} \\ 
&=&Tr\sqrt{\sum_{i=1}^l p_i  \sqrt{ \sigma} \rho_i \sqrt{\sigma}} \\ \nonumber
&\leq& \sum_{i=1}^l  \sqrt{p_i} \,Tr\sqrt{\sqrt{\sigma} \rho_i \sqrt{\sigma}}\\ \nonumber
&=& \sum_{i=1}^l  \sqrt{p_i}\,F(\sigma,\rho_i) \nonumber
\een
The second inequality in (\ref{eqn:graaf_inq}) can be rewritten as 
\be
F(\sigma,\tau)^2 \leq  1-{1\over 4}\|\sigma-\tau\|^2, 
\ee
which combined with the hypothesis $\|\rho_i-\sigma\|\geq 2-\epsilon$ leads to 
\be
F(\sigma,\rho_i)\leq  
\sqrt{ 1-{1 \over 4}(2-\epsilon)^2} =  \sqrt{\ep - \frac{\ep^2}4}  \leq  \sqrt{\ep} . 
\ee


Inserting this bound into (\ref{eqn:fid}) and using Cauchy-Schwarz inequality yields
\be
F(\sigma,\rho)\leq \sum_{i=1}^l  \sqrt{p_i}\sqrt{\ep} \leq \sqrt{l\ep} .
\ee
We are now in a position to appeal to the first of the Fuchs-van de Graaf inequalities \eqref{eqn:graaf_inq} to obtain 
\begin{flalign}
{1\over2}\|\rho-\sigma\|\geq 1-F(\sigma,\rho)\geq 1- \sqrt{l\epsilon} 
\end{flalign}
or $\|\rho-\sigma\|\geq 2- 2 \sqrt{l\epsilon}$, as needed. 

The dependence of the bound in \eqref{eq:szarek} on $\ep$ (and presumably on $l$) 
can not be significantly improved. To put this in a perspective, let us state an analoguous 
result for classical states, i.e., probability densities (non-negative functions with unit integral). 

\smallskip \noindent {\sl Let  $g_i, h$ be probability densities satisfying  $\|g_i -h\|_1=\int |g_i-h| \geq 2 - \ep$ for  $i=1,\ldots,l$. Then, for any weights $\{p_i\}_{i=1}^l$,
\be \label{eq:classical}
 \|\sum_{i=1}^l p_i g_i - h\|_1 \geq 2 - l\ep
\ee
and the inequality is sharp.}

\smallskip \noindent Indeed, since 
 for $u,v\geq 0$, $|u-v| = u+v -2 \min\{u,v\}$, the condition
$\|g_i-h\|_1 =\int g_i+h - 2 \min\,\{g_i,h\}  \geq 2-\ep$ translates to $\int \min\,\{g_i,h\}  \leq \ep/2$. Accordingly,  if $g = \sum_{i=1}^l p_i g_i$, then 
\be
\min\,\{g,h\}  \leq \sum_{i=1}^l \min\,\{g_i,h\} 
\ee
and so $\int \min\,\{g,h\}  \leq l \ep/2$, which is again equivalent to $\|f-g\|_1  \geq 2-l\ep$.

While the ``threshold for significance'' in the bounds in \eqref{eq:szarek} 
and \eqref{eq:classical} is roughly the same ($l\ep \ll 1$), 
the dependence on $l\ep$ as that quantity goes to $0$ is different. What is interesting 
is that this difference between the classical and quantum settings is real and not just an artifact of the argument. What follows is an example showing that the $\mathcal{O}(\sqrt{\epsilon})$ dependence in  \eqref{eq:szarek} is optimal. We will focus on the case $l=2$. 

To simplify the exposition, let us first reformulate the problem by considering a slightly more general question: {\em What is the optimal function $\ep \mapsto \phi(\ep)$ such that whenever 
$\rho_1,\rho_2,\sigma$ are positive semi-definite matrices whose trace is at most $1$ and such that 
$\tr \rho_i+\tr \sigma - \|\rho_i -\sigma\| \leq  \ep$ for  $i=1,2$, then $\tr \rho+\tr \sigma-\|\rho- \sigma\| \leq \phi(\ep)$ 
for any convex combination $\rho=p \rho_1 +(1-p) \rho_2$?} The point is  that the optimal function $\phi$ for this relaxed problem is the same as for the original problem 
when all the traces are required to be {\em equal to} $1$ at the cost of increasing the dimension by $2$. Indeed,  if $\rho_i,\sigma$ are as above, 
we may define states $\tilde{\rho}_i,\tilde{\sigma}$ by 
\be\tilde{\rho}_i= \left[
\begin{array}{ccc}
\rho_i & 0&0  \\
 0 & 1-\tr \rho_i & 0\\
 0& 0&0
\end{array}
\right], \ 
 \tilde{\sigma} =  \left[
\begin{array}{ccc}
 \sigma & 0  &0 \\
  0& 0&0 \\
0 & 0& 1-\tr \sigma
\end{array}
\right]
\ee
It is then easy to see that 
$2-\|\tilde{\rho}_i-\tilde{\sigma}\| = \tr \rho_i+\tr \sigma -\|\rho_i-\sigma\| $, 
and similarly for $\tilde{\rho}=p \tilde{\rho}_1 +(1-p) \tilde{\rho}_2$. 

With this reformulation, it is enough to look at $2\times 2$ matrices 
and $p_1=\frac 12$. 
Given  $r \in [0,1]$,  consider
\be
{\sigma}= \left[
\begin{array}{cc}
0 & 0  \\
 0 & r 
\end{array}
\right], \ 
{\rho}_i =  \left[
\begin{array}{ccc}
1-r & \pm \sqrt{r(1-r)}  \\
 \pm \sqrt{r(1-r)} & r 
 \end{array}
\right] ,
\ee
where $i=1$ corresponds to the plus sign and $i=2$ to the minus.  
One directly checks that 
\be 
\tr \rho_i+\tr \sigma -\|\rho_i-\sigma\|  = 1+r - \sqrt{1+2r-3r^2}  
\ee
On the other hand, if $\rho = \frac 12(\rho_1+\rho_2)$, then 
\be 
\tr \rho+\tr \sigma -\|\rho-\sigma\|  = 2r .
\ee 
In our setting, this means that if $\ep := 1+r - \sqrt{1+2r-3r^2} $
(which covers all possible values $\ep \in [0,2]$ as $r$ varies over $[0,1]$), then 
$\phi(\ep) \geq 2r$.  Since 
$(2r)^2 \geq  2\big(1+r - \sqrt{1+2r-3r^2}\big) = 2\ep $  for $r\in [0,1]$, 
this shows that $\phi(\ep) \geq \sqrt{2\ep}$. In other words, for $l=2$
one can not have a lower bound  in  \eqref{eq:szarek} 
that is better than $2- \sqrt{2\ep}$. 

While this example does not directly address the case $l>2$, 
we know that -- already in the classical setting -- 
one can not have a nontrivial bound if $l\ep$ is not small enough, 
and so the dependence of the bound  in  \eqref{eq:szarek} on $l$  
can not be too far from optimal.

\section{FOD in QM} \label{app:FODinQM}

\begin{prop}
The following quantity $c_0$ satisfies \eqref{eq:cond}
\be \label{classicbound}
c_0=\inf_{\xi,\xi',\{X_r\}} \max_{r,i,j} \min\{ p_i \tr(X_r \rho_i), q_j\tr(X_r \sigma_j)\}
\ee
where the infimum is taken over ensembles $\xi=\{(p_i,\rho_i)\}_{i=1}^{|y|}$,
$\xi'=\{(q_j,\sigma_j)\}_{j=1}^{|y'|}$ satisfying 
\be 
\sum_ip_i \rho_i = \sum_j q_j \sigma_j
\ee
and over all POVMs $\{X_r\}_{r=1}^k$, i.e., sets of operators satisfying $\sum_r X_r=I$, $X_r\geq 0$,
with $k=\max\{|x_1|,\ldots,|x_n|\}$.
\end{prop}

{\bf Proof.} 
We rewrite joint probabilities in terms of conditional probabilities. 
Clearly, for a given ensemble and a fixed input $x_n$, we can find two numbers $(a_0,b_0)$ such that,
\begin{flalign} \nonumber
&p_{ab|XY}(a_0,b_0):= &\\ \nonumber
&\max_{a,b,b'} \min \{ p(b|y) p(a^{(n)}|x_n,b), p(b'|y') p(a^{(n)}|x_n,b')\}&\\ \nonumber
&\ge \max_{r, i, j} \min \{p_i tr(X_r\rho_i), q_j tr(X_r\sigma_j)\}&
\end{flalign}
where, $p_i=p(b|y)$ and $q_j=p(b'|y)$ and $tr(X_r\rho_i)$ and $tr(X_r\sigma_j)$ are conditional probabilities. And we consider POVMs $\{X_r\}_{r=1}^k$ with $k=\max\{|x_1|,\ldots,|x_n|\}$.
Taking infimum over all ensembles $\xi$, $\xi'$ and inputs $\{X_r\}$, gives us RHS$=c_0$. Hence the proposition.
$\blacksquare$

\begin{lemma}
Suppose that $\|\rho-\sigma\|\leq 2- 2 k \ep$. 
Then for any POVM $\{X_r\}_{r=1}^k$ there exists an outcome $r_0$ such that 
\be
\tr (X_{r_0} \rho) \geq \ep, \quad \mbox{and} \quad \tr (X_{r_0} \sigma) \geq \ep
\label{eq:error}
\ee
\label{lem:error}
\end{lemma}

{\bf Proof.}
One shows that if, on the contrary, 
for all $r$ we have either $\tr (\rho X_r)\leq \ep$  or $\tr(\rho X_r)\leq \ep$,
then 
\be
p_e\leq \frac12 \ep k,
\ee
where $k$ is the number of outcomes of the POVM,
and $p_e$ is probability of error in distinguishing $\rho$ versus $\sigma$ 
with equal apriori probabilities given by  Helstrom relation
\be
p_e(\rho,\sigma) = \frac12 - \frac14 \|\rho-\sigma\|.
\ee
To prove it, let us define two sets:
$I_\rho=\{r: \tr (\sigma X_r)\leq \ep \}$ and $I_\sigma= I\setminus I_\rho\}$ 
where $I$ is the set of all indices $r$. By the above assumption,
for all $r\in I_\sigma $ we have $\tr (\rho X_r)\leq \ep$. 
Our decision scheme will be now: if $r\in I_\rho$ 
then the state is $\rho$, otherwise it is $\sigma$. 
With this decision scheme we have 
\ben
p_e\leq \frac12 \tr(\sum_{r\in I_\rho} X_r \sigma) +\frac12 \tr(\sum_{r\in I_\sigma} X_r \rho)
\nonumber \\
\leq \frac12 |I_\rho| \ep  + \frac12 |I_\sigma| \ep  = \frac12 k \ep.
\een
Thus, if $p_e\geq \frac12 \ep k$  then, there must exist such an outcome $r_0$
that both inequalities \eqref{eq:error} hold. $\blacksquare $

\begin{lemma}\label{lemma-x}
For two ensembles $\{(p_i,\rho_i)\}_{i=1}^{|y|}$,
$\{(q_j,\sigma_j)\}_{j=1}^{|y'|}$ satisfying 
\be
\|\sum_ip_i \rho_i - \sum_j q_j \sigma_j\|\leq x
\ee
there exist $i_0$ and $j_0$ such that 
\be
\|\rho_{i_0}-\sigma_{j_0}\|\leq 2 - \ep
\ee
where $\ep$ is solution of the following equation
\be
2 - 2\sqrt{l_1l_2\ep}=x
\ee
where $|y|=l_1$ and $|y'|=l_2$.
\label{lem:iteracja}
\end{lemma}

{\bf Proof.} Let us first show it for $|y|=|y'|=2$. 
Let us assume on the contrary, that for all pairs $(i,j)$ of indices $\|\rho_i - \sigma_j\|\geq 2-\ep$. 
Then together with theorem \ref{thm:szarek} imply that 
\be
\|p_1 \rho_1+p_2 \rho_2 -q_1 \sigma_1-q_2 \sigma_2\| \geq 2- 4\sqrt{\ep}
\ee
%

However, since the two ensembles give rise to the same density matrix, we have 
\be
2- 4\sqrt{\ep}\leq x
\label{eq:ff}
\ee
This implies, that at least one of the pairs must satisfy 
\be
\|\rho_{i_0} - \sigma_{j_0}\| \leq 2-\tilde\ep 
\ee  
where $\tilde\ep$ is solution of \eqref{eq:ff}, if we put equality.

Let us now extend the proof to the more general case. By theorem \ref{thm:szarek} there is for all $i \in \{1,...,|y|\}$
\be
\|\rho_i - \sum_{j=1}^{|y'|} q_j  \sigma_i\| \geq 2 - 2\sqrt{|y'|\ep}
\ee
Applying the theorem \ref{thm:szarek} again, we obtain
\be
\|\sum_{i=1}^{|y|}p_i\rho_i - \sum_{j=1}^{|y'|} q_j  \sigma_i\| \geq 2 - 2\sqrt{|y'| |y| \ep}
\ee
hence by assumption
\be
x \geq 2 - 2\sqrt{|y'||y|\ep},
\ee
and we obtain the thesis.
$\blacksquare$

\begin{lemma}
Suppose we are given two ensembles 
\be
\ecal_1= \{p_i,\rho_i\}_{i=1}^{l_1},\quad \ecal_2= \{q_j,\sigma_j\}_{j=1}^{l_2}
\label{eq:ens2}
\ee
which give rise to the same density matrix. Let $p_i$ and $q_j$ 
be in decreasing order.  Let us denote:
\be
\delta_1= 1- \sum_{i=1}^{\tilde{l_1}} p_i,\quad \delta_2= 1- \sum_{j=1}^{\tilde{l_2}} q_j
\ee
We then consider new ensembles 
\be
\tilde\ecal_1= \{\tilde p_i,\rho_i\}_{i=1}^{\tilde{l_1}},\quad \tilde\ecal_2= \{\tilde q_j,\sigma_j\}_{j=1}^{\tilde{l_2}}
\label{eq:ens2}
\ee
where $\tilde p_i = p_i/(1-\delta_1)$,$\tilde q_j = q_j/(1-\delta_2)$.
Then the new ensembles satisfy
\be
\|\sum_{i=1}^{\tilde{l_1}}\tilde p_i \rho_i - \sum_{j=1}^{\tilde{l_2}}\tilde q_j \sigma_j\| \leq \frac{2 \max\{\delta_1,\delta_2\}}{1 - \min\{\delta_1,\delta_2\}}
\ee
\end{lemma} 
{\bf Proof.}
From triangle inequality
\begin{flalign} \label{eq:sum}
&0= \| \sum_{i=1}^{l_1} p_i \rho_i -\sum_{j=1}^{l_2} q_j \sigma_j \| \geq & \nonumber \\
&\| \sum_{i=1}^{\tilde{l_1}} p_i \rho_i -\sum_{j=1}^{\tilde{l_2}} q_j \rho_j \| - \delta_1 -\delta_2.&
\end{flalign}
We replaced $\| \sum_{i=\tilde{l_1}+1}^{l_1} p_i \rho_i -\sum_{j=\tilde{l_2}+1}^{l_2} q_j \rho_j \| \le \delta_1 -\delta_2.$
 Let $x_1=1-\delta_1,x_2=1-\delta_2$. Then 
\ben
&& \|  x_1 \sum_{i=1}^{\tilde{l_1}} \tilde p_i \rho_i -x_1 \sum_{j=1}^{\tilde{l_2}} \tilde q_j \sigma_j \|= \nonumber \\
&& =\|  x_1 \sum_{i=1}^{\tilde{l_1}} \tilde p_i \rho_i -x_2 \sum_{j=1}^{\tilde{l_2}} \tilde q_j \sigma_j 
+  (x_2-x_1) \sum_{j=1}^{\tilde{l_2}} \tilde q_j \sigma_j\| \nonumber \\
&&  \leq  \| \sum_{i=1}^{\tilde{l_1}} p_i \rho_i -\sum_{j=1}^{\tilde{l_2}} q_j \sigma_j \|+|\delta_1-\delta_2|
\een 
Using \eqref{eq:sum}  we finally get 
\be
\|\sum_{i=1}^{\tilde{l_1}} \tilde p_i \rho_i -\sum_{j=1}^{\tilde{l_2}} \tilde q_j \sigma_j \|
\leq \frac{\delta_1+\delta_2-|\delta_1-\delta_2|}{1-\delta_1}.
\ee
Using $a+b+|a-b|=\max\{a,b\}$ and noticing that the same estimate holds, if we exchange 
$\delta_1$ with $\delta_2$, we obtain the required estimate.

\section{FOD \& CF for $l=2$}
Here we calculate bounds of {\it FOD} and {\it CF} in the case when Bob has binary outcomes. By measuring y or y' Bob can create $\{p_i\rho_i\}^1_{i=0}$ or $\{q_j\sigma_j\}^1_{j=0}$ ensembles at Alice's site  respectively. Using lemma (\ref{lem:error}) we know that for all pairs of $\rho_i$ and $\sigma_j$
\begin{flalign} \nonumber
\exists \hspace{5mm} \epsilon_{ij} \ge 0, X_{r_0}, X_{r_1}, X_{r_2}, X_{r_3} \hspace{4mm} s.t.\\ \nonumber
tr(X_{r_0}\rho_0)\ge \epsilon_{00} , \hspace{3mm}{\text and} \hspace{3mm} tr(X_{r_0}\sigma_0)\ge \epsilon_{00} \\ \nonumber
tr(X_{r_1}\rho_0)\ge \epsilon_{01} , \hspace{3mm}{\text and} \hspace{3mm} tr(X_{r_1}\sigma_1)\ge \epsilon_{01} \\ \nonumber
tr(X_{r_2}\rho_1)\ge \epsilon_{10} , \hspace{3mm}{\text and} \hspace{3mm} tr(X_{r_2}\sigma_0)\ge \epsilon_{10} \\ \nonumber
tr(X_{r_3}\rho_1)\ge \epsilon_{11} , \hspace{3mm}{\text and} \hspace{3mm} tr(X_{r_3}\sigma_1)\ge \epsilon_{11}
\end{flalign}

And FOD and Classical fraction (CF) for a fixed ensemble are given by,
\begin{flalign} \nonumber
FOD=\frac{1}{2k}\max\{&\min\{p_0 \epsilon_{00},q_0 \epsilon_{00}\},\min\{p_1 \epsilon_{11}, q_1 \epsilon_{11}\}, &\\ \nonumber
&\min\{p_0 \epsilon_{01},q_1 \epsilon_{01}\},\min\{p_1 \epsilon_{10},q_0 \epsilon_{10}\}\}&
\end{flalign}
and
\begin{flalign} \nonumber
CF=\frac{1}{2k}\max\{&\min\{p_0 \epsilon_{00},q_0 \epsilon_{00}\}+\min\{p_1 \epsilon_{11}, q_1 \epsilon_{11}\}, &\\ \nonumber
&\min\{p_0 \epsilon_{01},q_1 \epsilon_{01}\}+\min\{p_1 \epsilon_{10},q_0 \epsilon_{10}\}\}&
\end{flalign}

W.l.g. we assume $p_0\le q_0\le q_1\le p_1$. Hence we can rewrite above expressions as
\be
FOD=\frac{1}{2k}\max\{p_0 \epsilon_{00}, q_1 \epsilon_{11}, p_0 \epsilon_{01}, q_0 \epsilon_{10}\}
\ee
and
\be
CF=\frac{1}{2k}\max\{p_0 \epsilon_{00}+q_1 \epsilon_{11}, p_0 \epsilon_{01}+q_0 \epsilon_{10}\}
\ee

To simplify calculations one can express $\epsilon_{ij}$ in terms of $p_i$ and $q_j$ using lemma(\ref{lem:cut}), i.e., $\epsilon_{ij}\ge 2(1-{\max\{p_{\tilde i}, q_{\tilde j}\} \over 1-\min\{p_{\tilde i}, q_{\tilde j}})$. Then all needed to be done is optimise over p's and q's to calculate {\it FOD} and {\it CF}.
 And hence, 
\begin{flalign} \nonumber
&\epsilon_{00}\ge 2(1-{\max \{p_1, q_1\} \over 1-\min \{p_1, q_1\}})\ge 0 \ge 2(1-{p_1\over q_0})&\\ \nonumber
&\epsilon_{01}\ge 2(1-{\max\{p_1, q_0\} \over 1-\min\{p_1, q_0\}})\ge 0 \ge 2(1-{p_1\over q_1})&\\ \nonumber
&\epsilon_{10}\ge 2(1-{\max\{p_0, q_1\} \over 1-\min\{p_0, q_1\}})\ge 2(1-{q_1\over p_1})&\\
&\epsilon_{11}\ge 2(1-{\max\{p_0, q_0\} \over 1-\min\{p_0, q_0\}})\ge 2(1-{q_0\over p_1})&
\end{flalign}

Using theorem \ref{thm:szarek} one can show that  $\ep_{th}:=\max_{ij}\ep_{ij}\ge 1/4$.  
Considering different cases of $\epsilon_{ij}$ being equal to $\ep_{th}$
\begin{align} \label{eq:eps_frac}
\epsilon_{00}\ge {1\over 4}&\rightarrow \max \{{p_0\over 4}, 2q_1(1-{q_0\over p_1}), 0, 2q_0(1-{q_1\over p_1})\}&\\ \nonumber
\epsilon_{01}\ge {1\over 4}&\rightarrow \max \{0, 2q_1(1-{q_0\over p_1}), {p_0\over 4}, 2q_0(1-{q_1\over p_1})\}&\\ \nonumber
\epsilon_{10}\ge {1\over 4}&\rightarrow \max \{0, 2q_1(1-{q_0\over p_1}), 0, {q_0\over 4})\}&\\ \nonumber
\epsilon_{11}\ge {1\over 4}&\rightarrow \max \{0, {q_1\over 4}, 0, 2q_0(1-{q_1\over p_1})\}&
\end{align}

Minimizing over all ensemble with the constrains, i.e., $p_0\le q_0 \in [0,1/2]$ 
\begin{align} \label{eq:cls_frac} \nonumber
&\epsilon_{00}\ge {1\over 4}\rightarrow&\\ \nonumber
&\min_{p_0\le q_0 \in [0,{1\over2}]}[\max \{{p_0\over 4}, 2q_1(1-{q_0\over p_1})\}] \ge {5-\sqrt17\over 8}=0.1096&\\ \nonumber
&\epsilon_{01}\ge {1\over 4}\rightarrow &\\ \nonumber
&\min_{p_0\le q_0 \in [0,{1\over2}]}[\max \{2q_1(1-{q_0\over p_1}), {p_0\over 4}\}] \ge {5-\sqrt17\over 8}=0.1096&\\ \nonumber
&\epsilon_{10}\ge {1\over 4}\rightarrow &\\ \nonumber
&\min_{p_0\le q_0 \in [0,{1\over2}]}[\max \{2q_1(1-{q_0\over p_1}), {q_0\over 4})\}\ge 0.1123&\\ \nonumber
&\epsilon_{11}\ge {1\over 4}\rightarrow &\\ \nonumber
&\min_{p_0\le q_0 \in [0,{1\over2}]}[\max \{{q_1\over 4}, 2q_0(1-{q_1\over p_1})\}\ge 0.1123&
\end{align}

Hence taking minimum value of the above four values gives us
\begin{equation}
FOD \ge {1\over2k}(0.1096)
\end{equation}
Similarly one can calculate CF which is for this case,
\begin{equation}
CF \ge {1\over2k}(0.1123)
\end{equation}
\vfill

\end{document}